\documentclass{IEEE_lsens}
\usepackage{textcomp}
\usepackage[noadjust]{cite}
\usepackage[T1]{fontenc} % optional enhanced font encoding
\usepackage{amsmath}
\usepackage[cmintegrals]{newtxmath}
\usepackage{bm}
\usepackage{array}
\usepackage{url}
\usepackage[dvipsnames]{xcolor}

\usepackage{graphicx}
\usepackage{subfig}
\usepackage{algorithm}
\usepackage{algorithmic}
\newcommand{\sign}{\text{sign}}

\ifCLASSINFOpdf
\else
\fi
\providecommand{\hypersetup}[1]{\relax}

\hypersetup{pdftitle={Detecting Anomaly in Chemical Sensors via L1-Kernels based Principal Component Analysis},%<!CHANGE!
pdfsubject={Typesetting},%<!CHANGE!
pdfauthor={Hongyi Pan},%<!CHANGE!
pdfkeywords={Anomalous sensor detection, 
Principal component analysis, $\ell_1$-kernel, multiplication-free method.}}%<^!CHANGE!}
\begin{document}

% \markboth{Vol.~1, No.~3, July~2017}{0000000}

% article subject line 
\IEEELSENSarticlesubject{Sensor Applications}

\title{Detecting Anomaly in Chemical Sensors via L1-Kernel based Principal Component Analysis}

\author{\IEEEauthorblockN{Hongyi Pan\IEEEauthorrefmark{1}, Diaa Badawi\IEEEauthorrefmark{1}, Ishaan Bassi\IEEEauthorrefmark{2}, Sule Ozev\IEEEauthorrefmark{2}, Ahmet Enis Cetin\IEEEauthorrefmark{1} \IEEEauthorieeemembermark{1}}% <-this % stops a space
\IEEEauthorblockA{\IEEEauthorrefmark{1}Department of Electrical and Computer Engineering,
University of Illinois Chicago, Chicago, IL, 60607, USA\\
\IEEEauthorrefmark{2}School of Electrical, Computer and Energy Engineering, Arizona State University, Tempe, AZ, 85281, USA\\
\IEEEauthorieeemembermark{1}Fellow, IEEE}%
% LSENS authors should provide a real e-mail address here.
\thanks{This work is supported by NSF grants 1739396, 1934915 (UIC) and 1739451 (ASU). Corresponding author is H. Pan (e-mail: hpan21@uic.edu).\protect\\
%(For the real e-mail address, see http://www.michaelshell.org/contact.html).
}% <-this % stops a space
%\thanks{Associate Editor: Alan Smithee.}%
%\thanks{Digital Object Identifier 10.1109/LSENS.2017.0000000}
}
%
% note the % following lines that end in } - 
% these prevent an unwanted space from occurring between the objects.
% i.e., if you had this:
% For what it is worth, this is a minor point as most people would not even
% notice if the said evil space somehow managed to creep in.

% Manuscript received line
%\IEEELSENSmanuscriptreceived{Manuscript received June 7, 2017;
%revised June 21, 2017; accepted July 6, 2017.
%Date of publication July 12, 2017; date of current version July 12, 2017.}

% for Sensors Letters papers, we must declare the abstract and index terms
\IEEEtitleabstractindextext{%
\begin{abstract}
We propose a kernel-PCA based method to detect anomaly in chemical sensors. We use temporal signals produced by chemical sensors to form vectors to perform the Principal Component Analysis (PCA). We estimate the kernel-covariance matrix of the sensor data and compute the eigenvector corresponding to the largest eigenvalue of the covariance matrix. The anomaly can be detected by comparing the difference between the actual sensor data and the reconstructed data from the dominant eigenvector. In this paper, we introduce a new multiplication-free kernel, which is related to the $\ell_1$-norm for the anomaly detection task. The $\ell_1$-kernel PCA is not only computationally efficient but also energy-efficient because it does not require any actual multiplications during the kernel covariance matrix computation. Our experimental results show that our kernel-PCA method achieves a higher area under curvature (AUC) score (0.7483) than the baseline regular PCA method (0.7366).
\end{abstract}

}

%\IEEEpubid{1949-307X \copyright\ 2017 IEEE. Personal use is permitted, but republication/redistribution requires IEEE permission.\\
%See \url{http://www.ieee.org/publications\_standards/publications/rights/index.html} for more information.}

\maketitle

%\section*{Acknowledgment}
% addcontentsline needed when using bookmark hyperlinking, etc.
%\addcontentsline{toc}{section}{Acknowledgment}
% enable scriptsize
%\scriptsize

%\textit{IEEE Sensors Letters} is rendered in scriptsize.

% put at least one blank line to end the scriptsize paragraph and
% then revert back to normalsize.
\small

\section{Introduction}
Chemical sensors are widely used to detect ammonia, methane, and other Volatile Organic Compounds (VOCs) ~\cite{vergara2012chemical,erden2010voc, badawi2020computationally}. 
%These chemicals need to be detected timely because they are not only carcinogenic but also the main contributors to the greenhouse effect. 
%Given the critical nature of chemical gas detectors, it is important to know they are working correctly at all times. 
The life and performance of chemical gas detection sensors can be affected by various factors, including temperature, humidity, other interfering chemical gases, physical factors etc. Anomalous sensors can produce drifting waveforms
and it is a fatal problem for reliable gas identification and concentration estimation~\cite{mittal2017single, egilmez2014spectral}. In this work, we determine anomalous sensors and sensor measurements in an array of uncalibrated sensors by using robust $\ell_1$-Principal Component Analysis (PCA) without using a reference time-series data.

PCA is used in anomalous sensor and sensor signal detection~\cite{huang2006network,chatzigiannakis2007diagnosing,erhan2021smart}. %The principal components can be employed used in a variety of applications. 
In this approach, the covariance matrix is constructed from a set of data vectors and the anomalous items (outliers) or vectors are found by using the reconstruction difference~\cite{yao2010online, sree2014anomaly}. The principal components of the data covariance matrix is computed  and the original data vectors are reconstructed using only the first few principal components. In general, the reconstructed data vector is similar to the original data vectors, and the reconstructed data items that are  different from the corresponding original items are considered to be anomalous. 

In this paper, we propose to use $\ell_1$-kernel PCA based on a multiplication-free kernel to detect anomaly in chemical sensors.
Although the conventional PCA, which is based on the $\ell_2$-norm has successfully solved many problems it is sensitive to outliers in the data because the effects of the outliers are over-amplified by the $\ell_2$-norm. Recently, it has been shown that $\ell_1$-norm based methods produce better results in practical problems compared to the $\ell_2$-norm-based methods in several real-world signal, image, and video processing problems. In particular, $\ell_1$-kernel PCA usually is more robust against outliers in data  compared to the  $\ell_2$-PCA \cite{pan2021robust}. In the ordinary $\ell_2$-PCA, the principal vector is calculated as the dominant eigenvector of the data covariance matrix, which itself is calculated using the standard outer products. In this paper, we propose the eigen-decomposition of the $\ell_1$-kernel covariance matrix obtained using a new vector product, which  induces the $\ell_1$-norm without performing any multiplications. Because of low computational complexity, the  $\ell_1$-kernel PCA can be implemented in edge devices directly connected to the chemical sensors.  

Related work includes the regular $\ell_2$-PCA, kernel-PCA methods \cite{scholkopf:kpca}, the recursive $\ell_1$-PCA~\cite{markopoulos2014optimal} and the efficient $\ell_1$-PCA via bit flipping~\cite{markopoulos2017efficient}. The recursive $\ell_1$-PCA~\cite{markopoulos2014optimal} and the efficient $\ell_1$-PCA via bit flipping~\cite{markopoulos2017efficient} returns the same result, while the former takes the exponential time and the latter takes the polynomial time. Both $\ell_1$-PCA methods~\cite{markopoulos2014optimal, markopoulos2017efficient} require some parameters to be properly adjusted and rely on recursive methods. On the contrary, the proposed $\ell_1$-kernel PCA approach does not need any hyper-parameters adjustments. Its implementation is as straightforward as the regular PCA because we construct a sample kernel-covariance matrix using the proposed $\ell_1$-kernel and obtain the eigenvalues and eigenvectors of the kernel covariance matrix to define the linear transformation instead of solving an optimization problem.

The rest of the paper is organized as follows: In Section \ref{secMethodology}, we formally introduce the  $\ell_1$-kernel PCA and describe its application in an anomalous chemical sensor detection task. In Section \ref{secNumerical}, we compare our method with the regular PCA, the recursive $\ell_1$-PCA~\cite{markopoulos2014optimal} and the efficient $\ell_1$-PCA via bit flipping~\cite{markopoulos2017efficient}. Finally, in Section \ref{secConclusion}, we draw our main conclusions.

\section{$L_1$-Kernel PCA}\label{secMethodology}
%\subsection{}
In our recent work~\cite{akbacs2017energy, pan2021robust}, we proposed a set of  kernel-based PCA methods related with the $\ell_1$-norm. These Mercer-type kernels are obtained from multiplication-free (MF) dot products.

Let $\mathbf{w} = [w_1 \cdots w_n]^T \in \mathbb{R}^{n\times 1}$ and $\mathbf{x} = [x_1 \cdots x_n]^T \in \mathbb{R}^{n\times 1}$  be two $n$-dimensional column vectors. 
%The standard Euclidean inner product is defined as
%\begin{align}
%\label{ell2innerproduct}
 %    \mathbf{w}^T \mathbf{x} \triangleq \sum_{i=1}^D w_i x_i
%\end{align}
Similar to the regular dot product of vectors we defined the multiplication-free (MF) vector product~\cite{pan2021robust}.
In vector data correlation operations, we use the following 
vector product:

\begin{align}
	\label{odotproduct}
	\mathbf{w}^T \odot \mathbf{x}  & \triangleq \sum_{i=1}^n \sign(w_i \times  x_i) \min(|w_i|, |x_i|),
\end{align}
which turns out to be a Mercer type kernel: $K(\mathbf{w}, \mathbf{x}) =\mathbf{w}^T \odot \mathbf{x} $.
%Notice that the only multiplication operations
In  Eq. (\ref{odotproduct}), $\sign(w_i \times x_i)$ can be computed without performing any actual multiplications and $\min$ operation can be implemented by subtraction and checking the sign of the result of the subtraction. For this reason, we call  Eq. (\ref{odotproduct}) a Multiplication-Free (MF) dot product. 
The dot product defined in Eq.~(\ref{odotproduct}) induces the $\ell_1$-norm as 
$
	\mathbf{x}^T \odot \mathbf{x} = \sum_{i=1}^n \min(|x_i|, |x_i|) = \|\mathbf{x}\|_1
$
and  it induces a Mercer-type kernel \cite{pan2021robust}.
%\begin{equation}
%	\mathbf{x}^T \odot_m \mathbf{x} = \sum_{i=1}^n \min(|x_i|, |x_i|) = \|\mathbf{x}\|_1
%\end{equation}

Suppose that we collect vectors of  sensor data and form a  dataset $\mathbf{X}=[\mathbf{x}_1 \ \mathbf{x}_2 \ ... \  \mathbf{x}_D]\in \mathbb{R}^{N\times D}$. The well-known $\ell_2$-PCA (regular PCA) method relies on the eigendecomposition of the sample covariance matrix $\mathbf{C} = \mathbf{X}^T\mathbf{X}$.  
%Vector dot products described above can be extended to matrix multiplications as follows: Let $\mathbf{W} \in \mathbb{R}^{n\times m}$ and $\mathbf{X} \in \mathbb{R}^{n\times p}$ be arbitrary matrices. 
Similarly, 
we estimate the kernel covariance matrix as follows:
\begin{align}
\label{l1samplecovariance}
\mathbf{A} = \mathbf{X}^T \odot \mathbf{X},
\end{align}where the matrix $\mathbf{A}$ is constructed using the dot products of the form $   \mathbf{x}_i^T \odot \mathbf{x}_j  $. %Note that the ordinary matrix product in Eq. (\ref{l2samplecovariance}) is replaced by the MF product in Eq. (\ref{l1samplecovariance}). 
As a result, the construction of the kernel-covariance matrix $\mathbf{A} $ is straightforward.
%In this way, our $\ell_1$-PCAs are achieved by replacing the standard covariance matrix with the MF covariance matrices. 
We name the kernel-PCA based on the vector product in Eq. (\ref{odotproduct}) as the $\ell_1$-kernel PCA.%, and the PCA of the matrix obtained using Eq. (\ref{eqn_mf_vector}) as the MF-PCA, respectively.
%the $\ell_1$-PCA based on MIN operation in Eq. (\ref{odotproduct}) as MIN-PCA, and the $\ell_1$-PCA based on MIN2 operation in Eq. (\ref{odotproduct2}) as MIN2-PCA.

\subsection{Anomaly Detection Using $L_1$-Kernel PCA}\label{secAnomaly Detection Using Kernel-PCA}

%We first consider the single sensor case. The sensor 3  produces impulsive spikes between 0 and 1000$s$ as shown in  Fig.\ref{fig: data1} (in orange). Similar to PCA-based denoising methods \cite{scholkopf:kpca, mike1999kernel} we form data vectors from neighboring temporal data windows and form the measurement data matrix $\mathbf{X}=[\mathbf{x}_1 \ \mathbf{x}_2 \ ... \  \mathbf{x}_L]\in \mathbb{R}^{N\times L}$ where $L$ is the number of data windows and we have $N$ samples in each window. We form the kernel covariance matrix $\mathbf{A}= \mathbf{X}^T \odot \mathbf{X}\in\mathbb{R}^{L\times L}$ and compute its eigenvalues and eigenvectors $\mathbf{v}_i\in\mathbb{R}^{L\times1}$. We normalize the sensor values to $[-1, 1]$ and subtract the mean before computing the  kernel covariance mmatrix $\mathbf{A}$. We reconstruct the data matrix $\hat{\mathbf{X}}=\mathbf{V}\mathbf{V}^T{\mathbf{X}}$ using the first $l$ eigenvectors  $\mathbf{V} =[{\bf v_1, ..., v}_l]$ where $\mathbf{v}_1$ is the eigenvector corresponding to the largest eigenvalue. After this step we compare the actual data vectors $\mathbf{x}_i$ with the reconstructed ones $\hat{\mathbf{x}}_i, i=1,2,...,N$. The vectors significantly different from the reconstructed vectors are considered to be anomalous. We let $l=1$ and observed that it is sufficient for anomaly detection.

\begin{figure}[htbp]
    \centering
    \subfloat[Sample data 1\label{fig: data1}]{\includegraphics[width=0.7\linewidth]{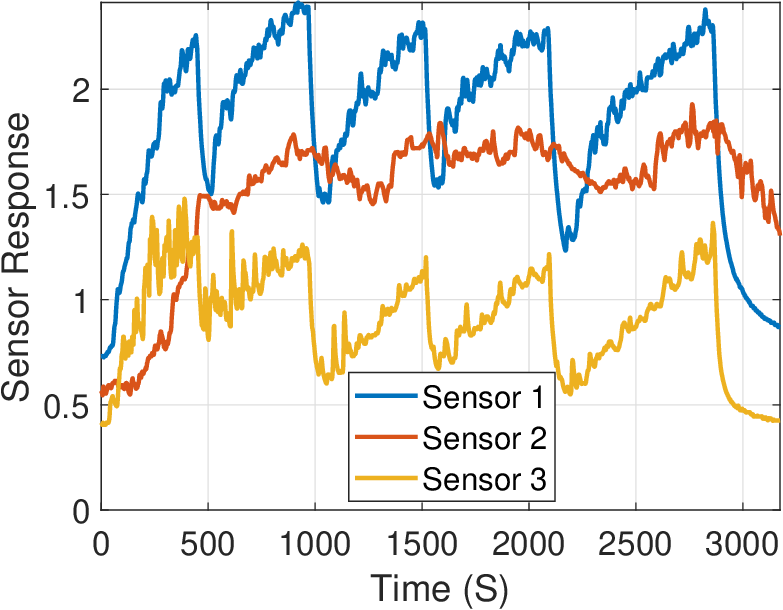}}\quad
    \subfloat[Sample data 2\label{fig: data2}]{\includegraphics[width=0.7\linewidth]{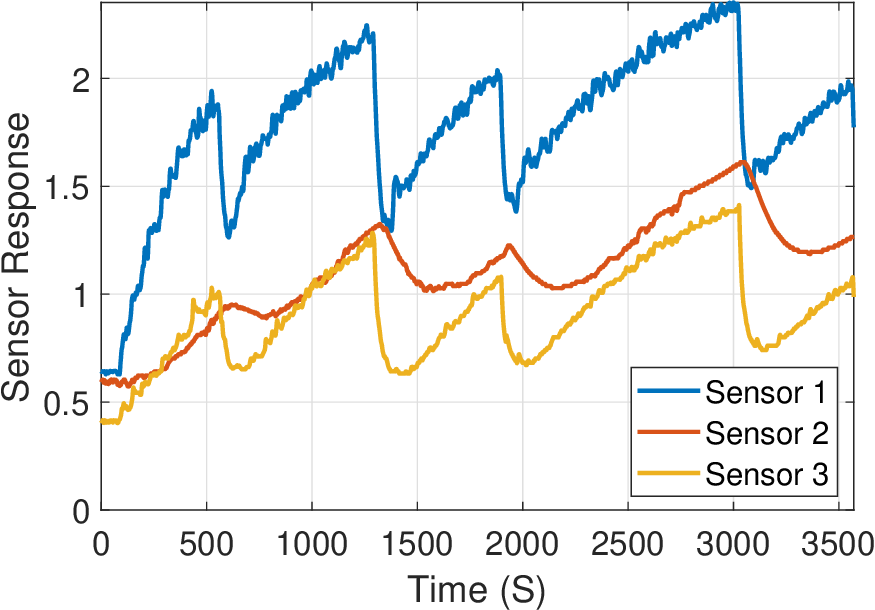}}
    \caption{Data produced by three ammonia sensors in two different ammonia gas recording experiments. The sensor 2 (in red) is the obstructed sensor. These data constitute our test set. }
    \label{fig: data}
\end{figure}

%In the second case w
We first assume that there are $D$ sensors and some of them are anomalous. Sensors are assumed to be close to each other and they produce correlated output waveforms as shown in Fig.~\ref{fig: data}. We have the measurement data $\mathbf{X}=[\mathbf{x}_1 \ \mathbf{x}_2 \ ... \  \mathbf{x}_D]\in \mathbb{R}^{N\times D}$. The measurement data is generated by normalizing the raw sensor measurement values to $[-1, 1]$ and subtracting the mean. We construct the covariance matrix $\mathbf{A}=\mathbf{X}^T\odot\mathbf{X}\in\mathbb{R}^{D\times D}$ and calculate its eigenvectors.
Let  $\mathbf{v}_1\in\mathbb{R}^{D\times1}$ be the dominant principal component vector. Finally, we reconstruct the data segment using the vector $\mathbf{v}_1$: $\hat{\mathbf{x}}_i = \mathbf{v}_1\mathbf{v}_1^T \mathbf{x}_i, i=1,2,..,D$ and compute the error vector ${\mathbf{x}}_i - \hat{\mathbf{x}}_i$. The sensor measurement in the $i$-th segment is assumed to be anomalous if the Cumulative Squared Difference (CSD) between $\hat{\mathbf{x}}_i$ and ${\mathbf{x}}_i$ is larger than a threshold. The threshold can be set as $T=\mu+ \alpha \sigma$, where $\mu$ and $\sigma$ are the mean and standard deviation of CSD values learned from a training data set. The parameter $\alpha$ is usually selected as 3 with the Gaussianity assumption of the CSD values. If a sensor produces successive anomalous measurement vectors it is considered to be anomalous.

In the second case, we assume that we only have a single sensor. For example, the sensor 3  produces impulsive spikes between 0 and 1200$s$ as shown in  Fig.\ref{fig: data1} (in orange). Similar to the PCA-based denoising methods \cite{scholkopf:kpca, mike1999kernel} we form data vectors from neighboring temporal data windows and form the measurement data matrix $\mathbf{X}=[\mathbf{x}_1 \ \mathbf{x}_2 \ ... \  \mathbf{x}_L]\in \mathbb{R}^{N\times L}$ where $L$ is the number of data windows and we have $N$ samples in each window. We form the kernel covariance matrix $\mathbf{A}= \mathbf{X}^T \odot \mathbf{X}\in\mathbb{R}^{L\times L}$ and compute its eigenvalues and eigenvectors $\mathbf{v}_i\in\mathbb{R}^{L\times1}$. We reconstruct the data matrix $\hat{\mathbf{X}}=\mathbf{V}\mathbf{V}^T{\mathbf{X}}$ using the first $l$ eigenvectors  $\mathbf{V} =[{\bf v_1, ..., v}_l]$ where $\mathbf{v}_1$ is the eigenvector corresponding to the largest eigenvalue. After this step we compare the actual data vectors $\mathbf{x}_i$ with the reconstructed ones $\hat{\mathbf{x}}_i, i=1,2,...,l$. The vectors significantly different from the reconstructed vectors are considered to be anomalous. We let $l=1$ and observed that it is sufficient for anomaly detection.

%, the sensor corresponding to the row is anomalous in this segment. %Anomaly detection algorithm using $\ell_1$-PCA is summarized in Algorithm~\ref{al: l1-PCA}.

%\begin{algorithm}[htbp]
%	\caption{Anomaly detection algorithm using $\ell_1$-PCA}
%	\begin{algorithmic}[1]
%		\renewcommand{\algorithmicrequire}{\textbf{Input:}}
%		\renewcommand{\algorithmicensure}{\textbf{Output:}}
%		\REQUIRE $\mathbf{X}=[\mathbf{x}_1 \ \mathbf{x}_2 \ ... \  \mathbf{x}_N]\in \mathbb{R}^{D\times N}$, anomaly threshould $T$.
%		\ENSURE Print anomalous sensors.
%		\FOR{$i=1, 2, ..., \frac{N}{L}-1$}
%		    \STATE $\mathbf{S}_i=[\mathbf{x}_{iL+1} \ \mathbf{x}_{iL+2} \ ... \  \mathbf{x}_{iL+L}]$,
%		    \STATE Normalize each row of $\mathbf{S}_i$ to $[-1, 1]$ to get $\bar{\mathbf{S}}_i$,
%		    \STATE $\mathbf{A}_i = \frac{1}{D}(\bar{\mathbf{S}}_i \oplus \bar{\mathbf{S}}_i^T)\in \mathbb{R}^{D\times D}$, where $\oplus\in\{\oplus_{mf},\odot,\odot_m\}$,
%		    \STATE $\mathbf{v}_i = \text{eigs}(\mathbf{A}_i, 1)\in \mathbb{R}^{D\times 1}$,
%		    \STATE $\hat{\mathbf{s}}_i=\mathbf{v}_i^T\bar{\mathbf{S}}_i\in\mathbb{R}^{1\times L}$
%		    \FOR{$j=1, 2, ..., D$}
%		        \STATE $CSD = \text{sum}((\hat{\mathbf{s}}_i-\bar{\mathbf{s}}_j)^2)$, where $\bar{\mathbf{s}}_j$ is the $j$-th row of $\bar{\mathbf{S}}_i$,
%		        \IF{$CSD > T$}
%		            \STATE \textbf{Print} "Sensor $j$ is anomalous in segment $i$."
%		        \ENDIF
%		    \ENDFOR
%		\ENDFOR
%		\RETURN $\mathbf{W}$.\\
%	\end{algorithmic}
%	\label{al: l1-PCA}
%\end{algorithm}

{\em Complexity Analysis:}
To calculate the covariance matrix $\mathbf{C}$, we perform $D^2N$ multiplications and $D^2(N-1)$ additions because we perform $N$ multiplications in each dot product. On the other hand, to calculate the $\ell_1$-kernel covariance matrix $\mathbf{A}$, we perform $D^2N$ sign operations, $D^2N$ min operations and $D^2(N-1)$ additions. According to Table I in \cite{nasrin2021enos}, a multiplication operation consumes about 4 times more energy compared to the MF-operations in compute-in-memory (CIM) implementation at 1 GHz operating frequency. In this letter, we have three sensors and we used an Arduino to collect data so energy efficiency will not be significant but it will be significant in a large network with its own hardware. Since the value of $D=3$ or $5$ is much smaller than the vector length $N=125$ or $224$, the eigenvector computation is negligible compared to the covariance matrix construction in this task. As a result, the $\ell_1$-kernel PCA is about 4 times more energy efficient in CIM implementation. It is also more energy efficient in many other processors because multiplications consume more energy than additions and subtractions. \cite{nasrin2021mf}.

{\em Contribution:} In our recent work~\cite{pan2021robust} we introduced the $\ell_1$-kernel PCA, while this paper introduces a novel method to employ the $\ell_1$-kernel PCA into the anomaly detection problem. Our experiments show that in the anomaly detection task, the $\ell_1$-kernel PCA produces better results than the regular PCA, the recursive $\ell_1$-PCA~\cite{markopoulos2014optimal} and the efficient $\ell_1$-PCA via bit flipping~\cite{markopoulos2017efficient} in our data set obtained from chemical sensors.

\begin{table*}[htbp]
    \centering
    \begin{tabular}{c|ccc|ccc|ccc}
    \hline\noalign{\smallskip}
    &\multicolumn{3}{c|}{\textbf{Regular PCA}}&\multicolumn{3}{c|}{\textbf{$L_1$-PCA via Bit Flipping~\cite{markopoulos2017efficient}}}&\multicolumn{3}{c}{\textbf{$L_1$-Kernel PCA}}\\
    \textbf{Segment (Time/S)}&\textbf{Sensor 1}&\textbf{Sensor 2}&\textbf{Sensor 3}&\textbf{Sensor 1}&\textbf{Sensor 2}&\textbf{Sensor 3}&\textbf{Sensor 1}&\textbf{Sensor 2}&\textbf{Sensor 3}\\
    \noalign{\smallskip}\hline\noalign{\smallskip}
0--249&7.54&9.44&2.20&8.12&9.01&2.37&8.01&9.08&2.21\\
250--499&\textbf{25.61}&18.48&9.23&\textbf{21.82}&\textbf{23.66}&8.71&19.54&\textbf{27.64}&7.24\\
500--749&11.81&12.19&17.13&16.46&9.23&16.57&14.06&10.96&16.59\\
750--999&0.71&\textbf{24.75}&1.08&0.76&\textbf{25.86}&0.85&0.75&\textbf{25.26}&0.90\\
1000--1249&3.03&9.11&11.88&2.93&9.04&12.08&3.14&8.70&12.35\\
1250--1499&4.57&7.02&1.30&6.03&6.48&1.89&4.64&6.92&1.38\\
1500--1749&5.71&\textbf{31.50}&2.79&10.60&\textbf{30.41}&1.16&7.84&\textbf{30.97}&1.73\\
1750--1999&3.77&\textbf{28.11}&3.52&1.31&\textbf{40.71}&0.87&3.32&\textbf{29.27}&3.01\\
2000--2249&1.29&14.74&2.04&1.13&15.34&1.83&1.46&14.61&2.13\\
2250--2499&2.09&\textbf{26.11}&3.55&1.94&\textbf{25.57}&5.40&1.98&\textbf{25.95}&3.88\\
2500--2749&5.08&6.08&7.39&5.13&5.53&8.61&4.95&6.32&7.37\\
2750--2999&1.09&10.67&1.87&1.05&10.91&1.77&1.17&10.66&1.86\\
3000--3249&0.98&12.89&3.51&0.92&13.22&3.30&0.99&12.94&3.51\\
         \noalign{\smallskip}\hline\noalign{\smallskip}
    \end{tabular}
    \caption{CSDs of the sample data 1 in Fig.~\ref{fig: data1} using regular PCA. Sensor 2 is obstructed. Values larger than threshold 21.60 are in bold.}
    \label{tab: CSD data1}
\end{table*}

\begin{table*}[htbp]
    \centering
    \begin{tabular}{c|ccc|ccc|ccc}
    \hline\noalign{\smallskip}
    &\multicolumn{3}{c|}{\textbf{Regular PCA}}&\multicolumn{3}{c|}{\textbf{$L_1$-PCA via Bit Flipping~\cite{markopoulos2017efficient}}}&\multicolumn{3}{c}{\textbf{$L_1$-Kernel PCA}}\\
    \textbf{Segment (Time/S)}&\textbf{Sensor 1}&\textbf{Sensor 2}&\textbf{Sensor 3}&\textbf{Sensor 1}&\textbf{Sensor 2}&\textbf{Sensor 3}&\textbf{Sensor 1}&\textbf{Sensor 2}&\textbf{Sensor 3}\\
    \noalign{\smallskip}\hline\noalign{\smallskip}
0--249&1.82&11.52&1.32&2.04&11.31&1.43&2.09&11.29&1.45\\
250--499&2.55&1.30&1.06&2.52&1.35&1.10&2.52&1.35&1.06\\
500--749&13.63&11.07&6.61&13.97&11.38&6.19&13.55&10.91&6.93\\
750--999&4.62&6.93&1.36&4.63&6.90&1.41&4.37&7.42&1.25\\
1000--1249&3.28&1.76&1.85&3.22&1.77&2.01&3.25&1.81&1.86\\
1250--1499&4.76&\textbf{50.56}&1.51&3.67&\textbf{52.57}&1.21&5.12&\textbf{50.12}&1.92\\
1500--1749&1.68&\textbf{28.23}&0.92&2.03&\textbf{27.83}&1.08&2.10&\textbf{27.71}&1.20\\
1750--1999&0.98&21.09&3.91&0.98&21.09&3.92&1.22&20.74&4.20\\
2000--2249&2.65&18.01&18.88&2.83&17.34&19.43&2.41&20.60&17.05\\
2250--2499&5.67&4.72&1.41&5.67&4.73&1.40&5.52&4.95&1.39\\
2500--2749&3.62&1.58&0.61&3.58&1.63&0.63&3.57&1.65&0.62\\
2750--2999&1.97&1.55&1.81&1.96&1.58&1.79&1.98&1.61&1.78\\
3000--3249&5.54&\textbf{50.49}&1.79&3.67&\textbf{54.51}&0.92&5.72&\textbf{50.23}&1.99\\
3250--3499&3.58&\textbf{37.25}&1.29&4.62&\textbf{35.78}&1.93&3.13&\textbf{38.00}&1.06\\
3500--3749&2.53&\textbf{35.25}&4.93&1.23&\textbf{40.35}&3.07&2.42&\textbf{35.53}&4.78\\
         \noalign{\smallskip}\hline\noalign{\smallskip}
    \end{tabular}
    \caption{CSDs of the sample data 2 in Fig.~\ref{fig: data2} using regular PCA. Sensor 2 is obstructed. Values larger than threshold 21.60 are in bold.}
    \label{tab: CSD data2}
\end{table*}

\section{Experimental Results}\label{secNumerical}
\begin{figure}[htbp]
    \centering
    \includegraphics[width=\linewidth]{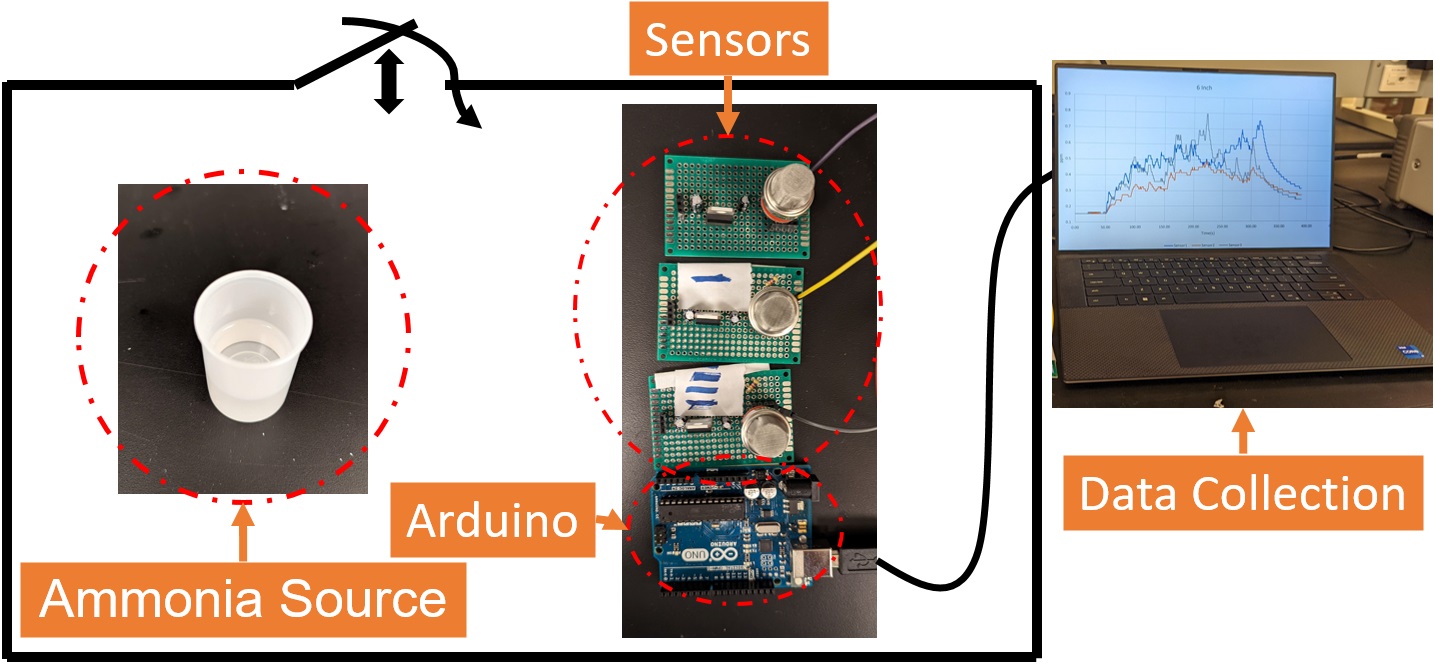} 
    \caption{Illustration of our experimental setup.}
    \label{fig:block_diagram}

\end{figure}

We collected data using  three ammonia MQ137 Tin oxide (SnO2) based sensors~\cite{watson1993tin}.
Sensors are connected to an Arduino Uno board, and the sampling rate is 2 samples per second. Sensors and a cylindrical ammonia source are placed in an airtight chamber. Sensors are pre-heated for 48 hours before collecting the data. When SnO2 is heated and exposed to the air, it reacts with the oxygen present in the air and form a layer of negative ion on the surface and reduce the surface conductivity~\cite{watson1993tin}. When ammonia vapor comes in contact with the surface, it combines with the oxide ion layer on the top and releases electrons for conduction. As a result, the conductivity of the surface increases. This change in surface resistance can be measured in the form of voltage. Our experimental setup is shown in Fig.~\ref{fig:block_diagram}.

{\em Multiple Sensor Anomaly Detection:}
The three sensors are placed close to each other and
%They should return similar ammonia concentrations. However,
one of the sensors (sensor 2) is obstructed with a cylindrical cover with multiple holes. The cover causes the sensor to react more slowly than the other sensors to the ammonia build-up and release. 
The obstruction level of the outlier sensor is adjusted in each trial to avoid over-fitting to one condition. Moreover, to generate a more realistic environment with varying levels of ammonia concentration, the chamber lid is opened at random intervals and with random duration. Opening and closing the lid is repeated multiple times to create different rise and fall responses. Sensor waveforms from two experiments are shown in Fig.~\ref{fig: data}.  % Impulsive sensor measurements are observed by other authors as well \cite{mittal2017single}. 

\begin{figure}[htbp]
    \centering
    \includegraphics[width=0.75\linewidth]{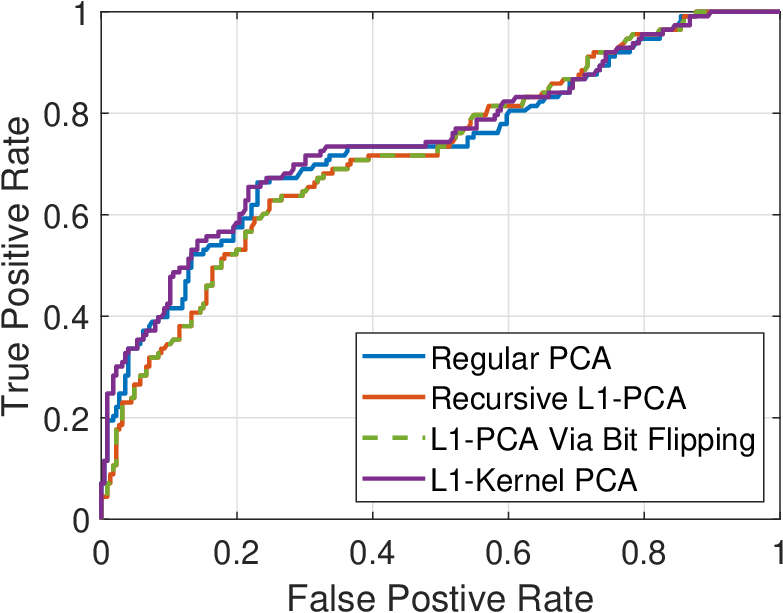}
    \caption{ROC of different PCAs on the sensor data. Recursive $\ell_1$-PCA~\cite{markopoulos2014optimal} and $\ell_1$-PCA via bit flipping~\cite{markopoulos2017efficient} return the same ROC because their eigenvectors are the same.}
    \label{fig: ROC}
\end{figure}

We apply the $\ell_1$-kernel PCA based anomaly detection method described in Section \ref{secAnomaly Detection Using Kernel-PCA} to the data obtained from the three sensors. We compared the proposed method with the regular $\ell_2$-PCA, the recursive $\ell_1$-PCA~\cite{markopoulos2014optimal} and the efficient $\ell_1$-PCA via bit flipping~\cite{markopoulos2017efficient}. The later two compute $
\mathbf{v}_{1} = \arg\max_{\mathbf{v}:\|\mathbf{v}\| = 1} \textstyle\sum_{i=1}^N |\mathbf{v}^T \mathbf{x}_i|$. Tolerance parameter of the recursive $\ell_1$-PCA method is set as $1\times10^{-8}$ as suggested by the authors~\cite{markopoulos2014optimal}.
We plot the Receiver Operating Characteristic (ROC) curve in Fig.~\ref{fig: ROC} and compute the Area Under Curve (AUC) scores for each method. As shown in Table~\ref{tab: AUC} states, the $\ell_1$-kernel PCA provides the highest AUC score, and the recursive $\ell_1$-PCA~\cite{markopoulos2014optimal}  provides the lowest AUC score in this case. This is probably due to the  non-convex optimization method that they use to compute the principle vector, and the method requires a suitable tolerance parameter. 
On the other hand, the $\ell_1$-kernel PCA does not need any tolerance parameters and the  eigenvector computations are equivalent to the computational load of the regular PCA.

\begin{table}[htbp]
    \centering
    \begin{tabular}{ccc}
    \hline\noalign{\smallskip}
    \textbf{Method}&\textbf{AUC score}&\textbf{AUC Increment}  \\
    \noalign{\smallskip}\hline\noalign{\smallskip}
         Regular PCA (baseline)&0.7366&-\\
         Recursive $\ell_1$-PCA~\cite{markopoulos2014optimal}&0.7203&-0.0163\\
         $L_1$-PCA via bit flipping~\cite{markopoulos2017efficient}&0.7203&-0.0163\\
         %MF-PCA&0.7441&0.0075\\
         \textbf{$L_1$-kernel PCA}&\textbf{0.7483}&\textbf{0.0117}\\
         %MIN2-PCA&0.7436&0.0070\\
         \noalign{\smallskip}\hline\noalign{\smallskip}
    \end{tabular}

    \caption{Area under curvature (AUC) for various methods. The $\ell_1$-kernel PCA provides the best AUC score. %All of our three PCA methods provide higher AUC scores than the regular PCA.
    }
    \label{tab: AUC}
\end{table}

CSD values of the sensors' response in Fig.~\ref{fig: data1} using different PCAs are listed in Table~\ref{tab: CSD data1}, and CSD values of the sensors' response in Fig.~\ref{fig: data2} are listed in Table~\ref{tab: CSD data2}, respectively. The threshold values 21.63 for the regular PCA, 21.64 for the $\ell_1$-PCA via bit flipping~\cite{markopoulos2017efficient} and 21.61 for the $\ell_1$-kernel PCA are computed from the normal cases obtained from other experiments. We have different threshold values for the regular PCA and the $\ell_1$-kernel PCA because they have different principle eigenvectors. Even if we use the same threshold (21.60) the classification results in Tables~\ref{tab: CSD data1} and~\ref{tab: CSD data2} will not change.
%We obtained the threshold value using three-sigma rule:  $Th=\mu+3\sigma$, where $\mu$ and $\sigma$ are the mean and standard deviation of the normal cases obtained from other experiments. 

%The AUC scores are relatively low in Table 1 because Sensor 2 does not always exhibit anomalous behavior.  In general,  its response increases due to ammonia gas exposure but not decrease as fast as the other sensors when there is no gas as shown in Fig. 1. 
%The regular PCA detects the anomalous behavior of Sensor 2 in 9 out of 28 data segments. On the other hand the $\ell_1$-kernel PCA detects the anomalous behavior of Sensor 2 in 10 out of 28 data segments. Moreover, the regular PCA produces a false alarm in the second data segment  (250 - 499 seconds) of Sensor 1 as shown in Table ~\ref{tab: regular PCA result2}.
%In Tables~\ref{tab: regular PCA result} and~\ref{tab: MIN-PCA result} the two methods report sensor 2 is anomalous at the same segment. However, in Table~\ref{tab: regular PCA result2}, there is a false-alarm case by the regular PCA at time segment 2000--2249, while in Table~\ref{tab: MIN-PCA result2} there is no false-alarm case by the MIN-PCA. Moreover, note that the threshold value of the regular PCA is higher in each case, so if we use the same threshold value 34.21 in sample data 1, the regular PCA will return a false positive case (sensor 3 at the third segment with the CSD 34.30). 

Sensor 2 does not always exhibit anomalous behavior.  In general,  its response increases due to ammonia gas exposure but not decrease as fast as the other sensors when there is no gas as shown in Fig.~\ref{fig: data}. The regular PCA detects the anomalous behavior of Sensor 2 in 9 out of 28 data segments. On the other hand, the $\ell_1$-PCA via bit flipping and our $\ell_1$-kernel PCA detect the anomalous behavior of Sensor 2 in 10 out of 28 data segments. Moreover, the regular PCA and the $\ell_1$-PCA via bit flipping produce a false alarm in the second data segment  (250 - 499 seconds) of Sensor 1 as shown in Table ~\ref{tab: CSD data2}, while our $\ell_1$-kernel PCA avoids this false alarm case. In conclusion, $\ell_1$-kernel PCA produces better results than the regular PCA, the recursive $\ell_1$-PCA~\cite{markopoulos2014optimal} and $\ell_1$-PCA via bit flipping~\cite{markopoulos2017efficient}.

{\em Anomaly Detection Using a Single Sensor}:
During the first three ammonia gas exposures, the sensor 3 positively responds but it also produces spikes up to 1200s as shown in Fig.~\ref{fig: data1}. Multisensor PCA-based anomaly detection cannot detect this behavior because only the sensor 1 works properly before 1200s. However, we compare the current sensor window of Sensor 2 with its neighboring data windows we can identify the anomalous behavior. We used $L=5$ data segments to construct the $5 \times 5$ covariance and $\ell_1$-kernel covariance matrices. In each data segment we have $N=224$ measurement. We used only the dominant eigenvector to estimate the data segments.

\begin{figure}[htbp]
    \centering
    \subfloat[Segment 1]{\includegraphics[width=0.5\linewidth]{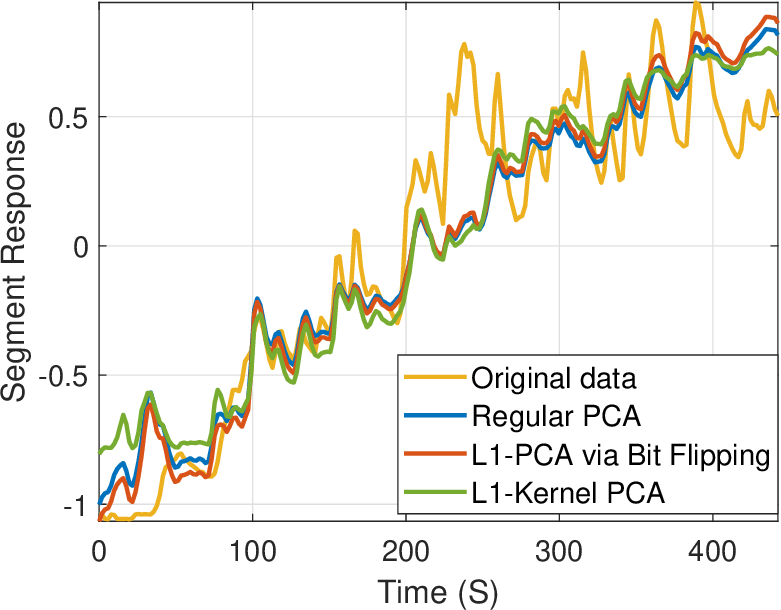}}
    \subfloat[Segment 2]{\includegraphics[width=0.5\linewidth]{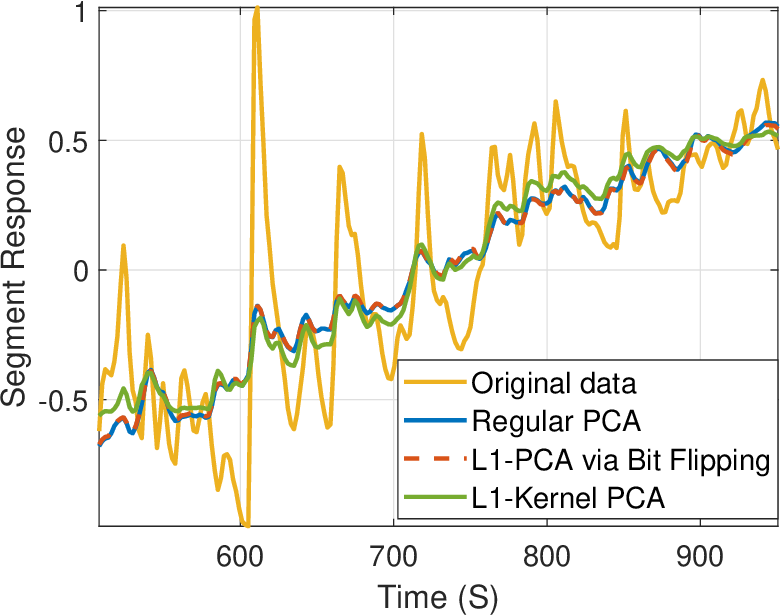}}%
    %\subfloat[Segment 3]{\includegraphics[width=0.33\linewidth]{data2_s3s3.eps}}
    \caption{Reconstructed waveforms of Segments 1 and 2 in Table~\ref{tab: Single Sensor}. The reconstructed waveforms do not have spikes.}
    %Response of the sensor 3 to ammonia gas exposures in orange. The first three segments contains impulsive spikes. The regular PCA and the $\ell_1$-kernel PCA reconstructed waveforms do not have spikes.}
    \label{fig: data_single}
\end{figure}

%In this case, assume we only have a single sensor (Sensor 3). The system will figure out if there exist impulsive spikes in the raising segment. As we mentioned in Section~\ref{secAnomaly Detection Using Kernel-PCA}, sensor 3 produces impulsive spikes between 0 and 1000 seconds as shown in  Fig.\ref{fig: data1} (in orange). 
The regular PCA, the $\ell_1$-PCA via bit flipping, and the $\ell_1$-kernel PCA reconstructed waveforms do not have spikes and that is how we can identify the anomaly in sensor readings as shown in Fig.~\ref{fig: data_single}. Table~\ref{tab: Single Sensor} shows the CSD values of these methods and they correctly identified the anomolous segments.
%detect that the CSD values in segments 1 and 2 are much larger than others.  
\begin{table}[htbp]
    \centering
    \begin{tabular}{cccc}
    \hline\noalign{\smallskip}
    \textbf{Segment}&\textbf{Regular PCA}&\textbf{$L_1$-PCA via Bit Flipping}&\textbf{$L_1$-Kernel PCA}\\
    \noalign{\smallskip}\hline\noalign{\smallskip}
            1&\textbf{9.62}&\textbf{9.51}&\textbf{9.79}\\
            2&\textbf{13.59}&\textbf{13.64}&\textbf{13.55}\\
            3&\textbf{7.21}&\textbf{7.30}&\textbf{7.26}\\
            4&4.35&4.58&4.38\\
            5&3.48&3.62&3.65\\
         \noalign{\smallskip}\hline\noalign{\smallskip}
    \end{tabular}

    \caption{CSD values due to ammonia exposure of Sensor 3. The first three segments contain impulsive spikes. }%Threshold values are 5.50 for the regular PCA, \textcolor{red}{5.61 for $L_1$-PCA via Bit Flipping} and 5.64 for the $\ell_1$-kernel PCA.}.

    \label{tab: Single Sensor}
\end{table}

\section{Conclusion}\label{secConclusion}
In this paper, we presented a framework for detecting anomalous sensors and sensor measurements in a chemical sensory system using the  $\ell_1$-kernel PCA. We collected data from three commercial Tin Oxide (SnO2) sensors by exposing them to ammonia in an environment-controlled experiment. 
%We stimulate anomalous responses by blocking one of the sensors, resulting in different characteristics in its respective readings. Then, 
%We compare our three methods with the regular PCA and the recursive $\ell_1$-PCA using these data. In our experiment, all of our methods provide higher AUC scores than the regular PCA and the recursive $\ell_1$-PCA. The AUC score of the MIN-PCA is 0.0145 higher than the regular PCA.
The proposed $\ell_1$-kernel PCA is more robust than the regular PCA in our experiments. This is due to the fact that $\ell_1$-kernel PCA is related with the $\ell_1$-norm and it gives less emphasis to anomalous spikes in sensor measurements while computing the correlation matrix. The computational energy cost of the $\ell_1$-kernel PCA is much lower than the regular PCA on many processors. Because of low energy complexity, the  $\ell_1$-kernel PCA can be implemented in low-cost edge devices directly connected to the chemical sensors.

% Last page column equalization
%
% IEEE Sensors Letters does balance the columns on the last page.
% Can use:
% \IEEEtriggeratref{8}
% to trigger a \newpage just before the given reference number to
% balance the columns on the last page. Adjust the reference number
% as needed - this may need to be readjusted if the document is 
% modified later.
% Alternatively, you can also directly use something like
% \enlargethispage{-7in}
% on the last page instead of breaking at a specific reference number.

%\bibliographystyle{IEEEtran}
% Before submitting to IEEE Sensors Letters, manually copy in the
% resultant .bbl file contents in place of the \bibliographystyle and
% \bibliography lines here:
\bibliographystyle{unsrt}
\bibliography{main}

\end{document}